\documentclass[fleqn,usenatbib, letters]{mnras}

\usepackage{mathptmx}
\usepackage[T1]{fontenc}
\usepackage{ae,aecompl}

\usepackage{graphicx}	% Including figure files
\usepackage{amsmath}	% Advanced maths commands
\usepackage{amssymb}	% Extra maths symbols

\title[XUV observations of PDS 70 ]{A \textit{Swift} view of X-ray and UV radiation in the planet-forming T-Tauri system PDS 70}

\author[S. R. G. Joyce et al.]{
Simon R. G. Joyce,$^{1}$\thanks{E-mail: sj328@leicester.ac.uk (SRGJ)}
John P. Pye,$^{1}$
Jonathan D. Nichols,$^{1}$
Kim L. Page,$^{1}$\newauthor
Richard Alexander,$^{1}$
Manuel G{\"u}del,$^{2}$
Yanina Metodieva$^{2}$
\\
$^{1}$School of Physics and Astronomy, University of Leicester, University Road, Leicester, LE1 7RH, UK\\
$^{2}$University of Vienna, Dept. of Astrophysics, Turkenschanzstr. 17, 1180 Vienna, Austria\\
}

\date{Accepted XXX. Received YYY; in original form ZZZ}

\pubyear{2019}

\begin{document}
\label{firstpage}
\pagerange{\pageref{firstpage}--\pageref{lastpage}}
\maketitle

% Abstract of the paper
\begin{abstract}
PDS 70 is a $\sim$5 Myr old star with a gas and dust disc in which several proto-planets have been discovered. We present the first UV detection of the system along with X-ray observations taken with the \textit{Neil Gehrels Swift Observatory} satellite. PDS 70 has an X-ray flux of 3.4$\times 10^{-13}$ erg cm$^{-2}$ s$^{-1}$ in the 0.3-10.0 keV range, and UV flux (U band) of 3.5$\times 10^{-13}$ erg cm$^{-2}$ s$^{-1}$ .
At the distance of 113.4 pc determined from Gaia DR2 this gives luminosities of 5.2$\times 10^{29}$ erg s$^{-1}$ and 5.4$\times 10^{29}$ erg s$^{-1}$ respectively. The X-ray luminosity is consistent with coronal emission from a rapidly rotating star close to the log $\frac{L_{\mathrm{X}}}{L_{\mathrm{bol}}} \sim -3$ saturation limit. We find the UV luminosity is much lower than would be expected if the star were still accreting disc material and suggest that the observed UV emission is coronal in origin.
\end{abstract}

% Select between one and six entries from the list of approved keywords.
% Don't make up new ones.

%%%%%%%%%%%%%%%%%%%%%%%%%%%%%%%%%%%%%%%%%%%%%%%%%%

%%%%%%%%%%%%%%%%% BODY OF PAPER %%%%%%%%%%%%%%%%%%

\section{Introduction}

Newly formed stars are often surrounded by discs of left-over gas and dust which are the site of planet formation. PDS 70 is a nearby (113 pc) example of a young T-Tauri star with a disc inclined at 45-50 degrees \citep{Hashimoto_2012}, making it possible to resolve and image the disc with optical and IR observations (e.g. \citealt{  Riaud_2006, Christiaens_2019_MNRAS_b, Keppler_2018}). Understanding conditions in this, and other similar systems, provides a direct window on the process of planet formation. Discs evolve on relatively short time-scales, and observations of clusters at different ages have shown that almost all proto-planetary discs disperse and disappear within $\sim$ 10 Myr (\citealt{Zuckerman_1995, Haisch_Lada_Lada_2001, Mamajek_2009, Ribas_2015}). 

There are several processes potentially contributing to the rapid disc evolution. The first stage of disc evolution is accretion onto the central star. In later stages, stellar X-ray and UV radiation (XUV) drives photo-evaporation which causes the central region to become cleared of dust and gas, eventually dispersing the entire disc (see reviews by \citealt{Ercolano_Pascucci_2017, Gorti_2016, Alexander_2014}). Planet formation is also expected to redistribute material in the disc and is capable of clearing gaps (e.g. \citealt{Kley_Nelson_2012, Zhu_2011}). It is likely that in most cases, a combination of these processes comes in to play, and their relative importance changes as the disc structure evolves.

Accurate XUV measurements are required as input to models of disc evolution which can begin to untangle the effects of photo-evaporation and ongoing planet formation on the disc (e.g. \citealt{Pascucci_2014, Gorti_Hollenbach_2009, Skinner_Gudel_2013, Alexander_2006}). It is also informative to estimate the long-term effects of this radiation on planets and their atmospheres which could go some way to explaining the large variety of planetary types which continue to be discovered.
 
Single stars below $\sim$ 1.25 M$_{\odot}$ are at their most active during the first $\sim$1 Gyr after formation (e.g. \citealt{Jackson_Davis_Wheatley_2012}), when the X-ray luminosity is saturated at log $L_{\textrm{x}}/L_{\textrm{bol}} \sim -3$. XUV radiation is associated with magnetic activity which is strongest for rapidly rotating stars (e.g. \citealt{Pallavicini_1981, Pizzolato_2003}). Activity and XUV luminosity decline as angular momentum is lost to the stellar wind, causing the rotation rate to slow down. High levels of radiation could have a significant effect on planetary atmospheres. It is estimated that for most of the Trappist-1 planets, the stellar XUV radiation would be sufficient to evaporate the equivalent of several Earth oceans per Gyr \citep{Wheatley_2017}, which does not bode well for habitability. The effect is especially significant for late K and M-dwarfs which maintain high activity levels for much longer than earlier spectral types. Planets within the habitable zones of late-type stars are therefore subjected to long-term erosion.

In this context PDS 70 presents an ideal opportunity to observe the effects of stellar XUV radiation on a planet-forming disc. PDS 70 was first identified as a T-Tauri star with a circumstellar disc using IR coronograph observations \citep{Riaud_2006}. The clearing of PDS 70's disc seems to be proceeding at different rates depending on the dust particle size. Small dust ($\mu$m) has been cleared to a radius of 53 AU whereas the gap observed in big dust ($\sim$mm) is larger, extending out to 65 AU (From \citealt{Hashimoto_2015} but recalculated using Gaia DR2 distance of 113.4 pc, see \citealt{Christiaens_2019_MNRAS_b}). It now appears that this is a gap rather than the entire central region being clear. Coronographic observations were unable to detect the inner region of the disc, but NIR imaging has since identified an inner edge to the gap at 17 AU \citep{Keppler_2018}.

One possible cause of the gap and dust stratification could be the action of proto-planets. The discovery of a planet in orbit at 22 AU gave support to this scenario \citep{Keppler_2018}. While many candidate proto-planets discovered by direct imaging have since been called in to question, several studies have confirmed the existence of PDS 70 b in NIR and H-alpha direct imaging (\citealt{Wagner_2018, Muller_2018, Christiaens_2019_MNRAS_b, Christiaens_2019_ApJ_a, Haffert_2019}). It has even been possible to detect the spectral signature of a circumplanetary disc around PDS 70 b and c \citep{Christiaens_2019_ApJ_a, Isella_2019} which may be forming moons \citep{Lunine_Stevenson_1982}. \cite{Haffert_2019} confirmed PDS 70 b is a 4-17 M$_{J}$ planet at 20.6 $\pm$ 1.2 AU and identified a second planet, PDS 70 c, at 34 $\pm$ 2 AU. The presence of a second planet is consistent with modelling which shows that a single planet could not produce the large (>15 AU) gap seen in PDS 70's disc (\citealt{Kley_Nelson_2012, Zhu_2011}). These discoveries confirm PDS 70 as a fascinating example of ongoing planet, and possibly moon formation, in the midst of an evolving proto-planetary disc.

With the \textit{Neil Gehrels Swift Observatory} \citep{Gehrels_2004} we have carried out the first simultaneous X-ray and UV (XUV) observations of PDS 70 in order to characterise the star's high-energy radiation and its potential effect on the disc and planets.

%

%Simultaneous observations with the UVOT and XRT instrument were carried out on four separate occasions over approximately 5 weeks. 

%===================================================
% Example table
\begin{table}

	\caption{Table of Swift observations of PDS 70 taken with the UVOT and XRT instruments simultaneously. Observations were taken on 5 separate occasions over a 5 week period. Wavelength ranges included in the filters (FWHM) are given in Table. \ref{Table:table_4_flux_lum} The photon counting (PC) mode of the XRT instrument detects the 0.3-10.0 keV energy range.}
	\label{Table:table_1_obs_list}

	\begin{tabular}{|c|c|c|c|}
	\hline 
	Date & Observation ID & Time & Filter/detector \\ 
	(dd-mm-yyyy) & (000114600..) &  &  \\ 
	\hline 
	07-07-2019 & 01/02 & 06:20:35 &  PC/uvw1/uvw2 \\

	13-07-2019 & 03/04/05 & 10:40:36 &  PC/uvw2 \\

	27-07-2019 & 06/07/08 & 01:13:24 &  PC/uvw1 \\

	28-07-2019 & 09/10/11 & 13:56:35 &  PC/U/uvw2  \\

	10-08-2019 & 12/13/14 & 04:47:34 &  PC/uvw2 \\ 
	      
	\hline 
	\end{tabular}

\end{table}

%===================================================

%===================================================
% Example table
\begin{table}
	\centering
	\caption{Stellar properties of PDS 70.}
	\label{Table:table_2_stellar_properties}

	\begin{tabular}{ccc} % four columns, alignment for each
		
		\hline
		Parameter & Value & Reference \\
		\hline
		RA (J2000) & 14:08:10.1545 & 1\\
		Dec (J2000) & -41:23:52.5766 & 1\\
		Spectral type & K7 & 2\\
		Age (Myr) & 5.4 $\pm$ 1.0 & 3\\
		Mass (M$_{\odot}$)  & 0.76 $\pm$ 0.02 & 3\\
		Radius (R$_{\odot}$)  & 1.26 $\pm$ 0.15 & 4\\
		Luminosity (L$_{\odot}$)  & 0.35 $\pm$ 0.09 & 4\\
		$T_{\rm eff}$ (K) & 3972 $\pm$ 36 & 4 \\
		Parallax (mas) & 8.8159 $\pm$ 0.0405 & 1\\
		Distance (pc) & 113.43 $\pm$ 0.52 & 1\\
		$M_{\rm disc}$ ($M_{\mathrm{Jup}}$) & 4.5 & 5\\
		$i_{\rm disc}$ (degrees) & 49.7 & 5\\
		\hline
		
	\end{tabular}
	
            1 = Gaia DR2, 2 = \cite{Gregorio_Hetem_2002}, 3 = \cite{Muller_2018}, 4 = \cite{Pecaut_Mamajek_2016}, 5 = \cite{Hashimoto_2015}

\end{table}

%===================================================

\section{Observations}

%=================================================

\begin{figure}
\includegraphics[width=84mm, height=33mm]{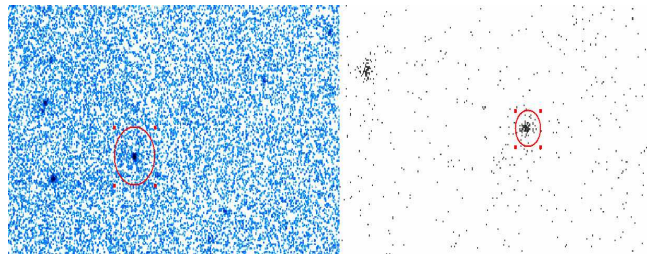}

\caption{\textbf{Left panel:} Swift UVOT observation using the uvw2 filter covering the 1599 - 2256 \AA\ wavelength range. The red circle was placed at the J2019 coordinates for PDS 70 from Gaia DR2 and has a 30 arcsec radius. This image is from a single observation 00011460002. 
\textbf{Right panel:} Same as left hand panel but for the XRT instrument in the 0.3 - 10 keV energy range. The data from all the observations taken in the photon-counting mode were combined to produce this image.}
\label{fig:source_detection}
\end{figure}

%=================================================

Swift observed PDS 70 on 5 occasions from 7th July 2019 to 10th August 2019 for approximately 2 ks per visit, giving a total of 10 ks exposure time. Swift observed simultaneously with the X-Ray Telescope (XRT) and the UV/Optical Telescope (UVOT). Table \ref{Table:table_1_obs_list} lists details of the individual observations. Basic parameters adopted for the PDS 70 system in this analysis are listed in Table~\ref{Table:table_2_stellar_properties}.

\subsection{X-ray and UV source detection}
The XRT instrument was used in photon counting mode which produces a 2-D image in the 0.3-10.0 keV energy range.  The source is only marginally detected in most of the individual observations. However, combining all the photon-counting mode exposures produces a clear source detection at the coordinates of PDS 70 as shown in Fig. \ref{fig:source_detection} right panel. This image includes the full 0.3-10.0 keV energy range. Images with only the 0.3-1.5 keV or the 1.5-10.0 keV range show that the source is  clearly detected in the low energy range, but very faint in the higher energy range. This is expected for a stellar coronal source where the majority of emission would be in the 0.5-2.0 keV range.

For the UV data the UVOT instrument has several filters covering slightly different wavelength ranges. The filter used depends on the current setting of the instrument at the time of the observation and therefore several different filters were used during the 5 observations. A UV source is clearly detected at the target coordinates in each individual observation, for example in the uvw2 filter covering the 1599-2256 \AA\ range as shown in Fig. \ref{fig:source_detection} left panel.

\subsection{Light-curves}

\begin{figure}
\includegraphics[width=84mm, height=42mm]{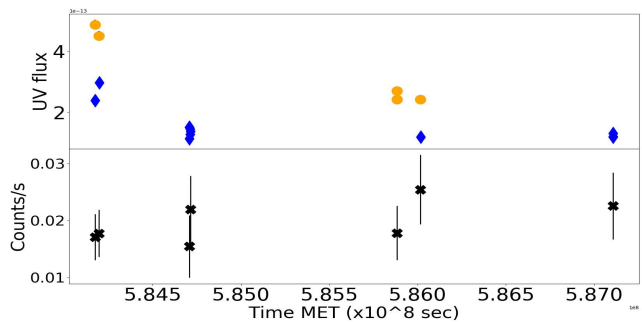}

\caption{\textbf{Top panel:} Swift UVOT flux ($\times 10^{-13}\ \mathrm{erg\ cm^{-2}\ s^{-1}}$) for each observation taken over 5 weeks. Observations were made using two filters. The uvw1 filter (orange circles) and uvw2 (blue diamonds).  
\textbf{Bottom panel:} The X-ray count-rate over the 5 observations from 07-07-2019 to 10-08-2019 in the 0.3-10 keV energy range. There is no significant change in the X-ray count-rate in contrast to the decline in the UV flux. }
\label{fig:light_curves}
\end{figure}

From the individual UVOT observations, the UV magnitude and flux were measured by selecting a 5 arcsec source region and a larger background region avoiding any visible sources. Flux measurements were made using the \textit{uvotmaghist} software. Flux measurements from individual observations are plotted in Fig \ref{fig:light_curves} (top panel) where orange circles indicate uvw1 (2253-2946 \AA)  and blue diamonds are uvw2 (1599-2256 \AA). They show that in both the uvw1 and uvw2 filters the first observation measured a higher flux. The following 4 observations all measured a slightly lower flux by about $1-2\times10^{-13}$ $\rm{erg\ cm^{-2}\ s^{-1}}$ and there was no significant variability between any of the 4 observations from 13-07-2019 to 10-08-2019.

X-ray count-rates were extracted from each observation using the automatic XRT data builder \citep{Evans_07}. 
The X-ray light-curve in Fig. \ref{fig:light_curves} lower panel shows no significant variability across all the observations, including the first one where there is a higher flux in the UV. Since there is very little variation in the X-ray light-curve and hardness-ratio, it appears the source of the X-rays is consistent across all the observations. Therefore it is possible to combine the X-ray data to produce a single spectrum which represents the source as shown in Fig. \ref{fig:spec_fit}.

%=================================================

%===================================================

\section{X-ray spectral analysis}
\begin{figure}
\includegraphics[width=84mm, height=58mm]{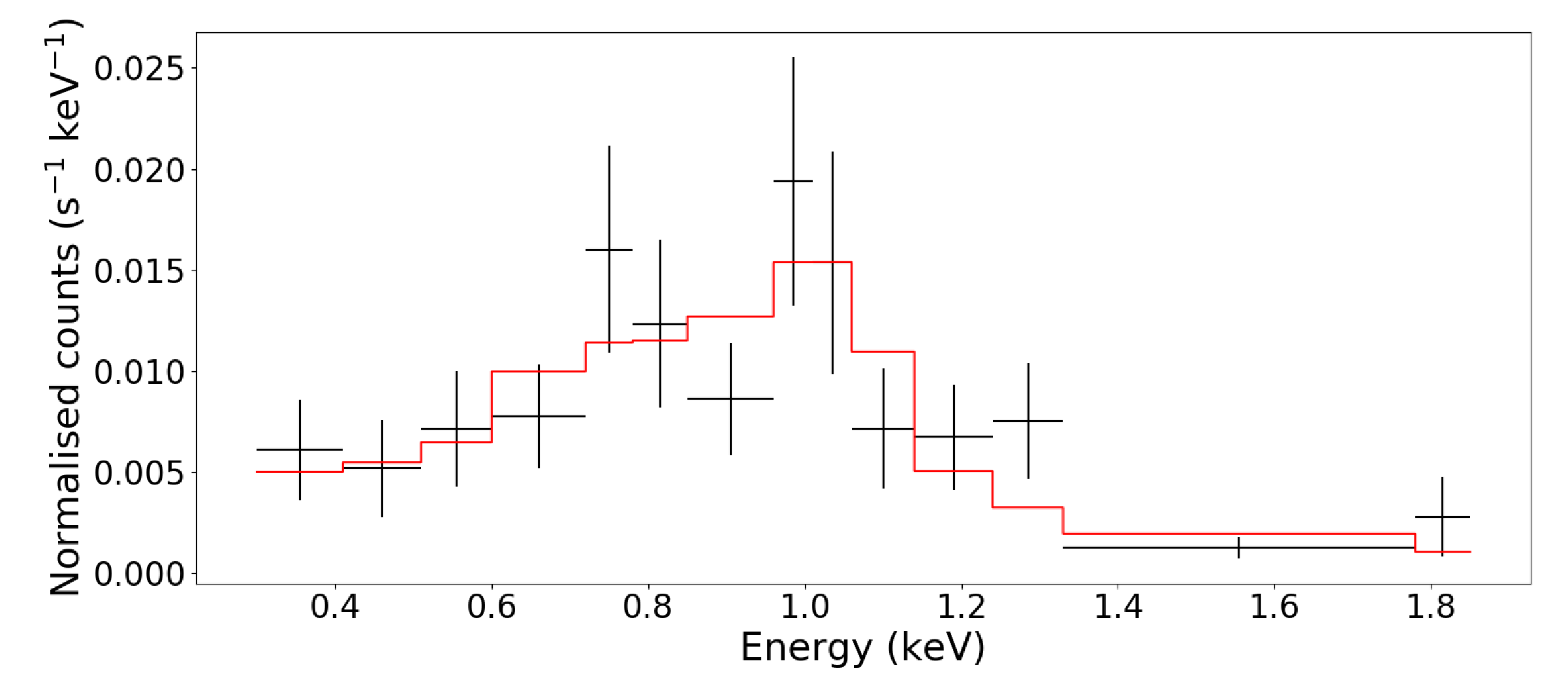}

\caption{The combined X-ray spectrum with a total exposure time of 10 ks. The spectrum is fitted with a 2-temperature \textsc{apec} model without absorption. Model parameters are given in Table \ref{Table:table_3_spec_fit_results}, column 4. }
\label{fig:spec_fit}
\end{figure}

Spectral files including ancillary response (ARF) and response matrix (RMF) were generated using the XRT automatic analysis tools \citep{Evans_09} and fitting was carried out using \textsc{xspec}. The spectrum has relatively few photons but shows a peak at $\sim $1 keV and very little flux above 2 keV which are characteristics of a stellar coronal spectrum \citep{Gudel_Naze_2009}. We therefore fit the spectrum with an \textsc{apec} plasma model \citep{Smith_2001} as shown by the red line in Fig. \ref{fig:spec_fit}.
The low number of counts in the spectrum requires the use of the C-statistic when fitting, rather than standard $\chi^{2}$. 

Single and 2-temperature models were compared, with results shown in Table \ref{Table:table_3_spec_fit_results}. Fitting with a 2-temperature model without absorption gave an improved fit with a reduced $\chi^{2}$ of 1.27 compared to the 1-temperature model's 1.42. The temperature of the cool plasma component is 0.31 keV which is consistent with the values found for the similar K5 type T-Tauri star LkCa 15 \citep{Skinner_Gudel_2013}. However the higher temperature plasma component is almost a factor 2 lower than the $\sim$2 keV found for LkCa 15. Adding an interstellar $N_{\mathrm{H}}$ absorption component to the 2-temperature model found negligible absorption and gave a slightly worse reduced $\chi^{2}$  while making very little difference to the flux derived from the models (See Table. \ref{Table:table_3_spec_fit_results}).

The 0.3-10 keV flux found for the 2-temperature model is 3.4$\times 10^{-13}$ erg cm$^{-2}$ s$^{-1}$ which, at the distance of PDS 70 (113.4 pc from Gaia DR2), gives a luminosity of 5.2$\times10^{29}$ $\rm{erg\ s^{-1}}$. The flux and luminosity measurements for both X-ray and UV are summarised in Table~\ref{Table:table_4_flux_lum}.

%=================================================

%===================================================
% Example table
\begin{table}
	\centering
	\caption{X-ray spectra fitting results.}
	\label{Table:table_3_spec_fit_results}
	\begin{tabular}{lccc} % four columns, alignment for each
		\hline
		Parameter &  &  & \\
		\hline
		Model & 1-T+tbabs &  2-T +tbabs  & 2-T no abs\\
		
		$kT_{1}$ (keV) & 0.14 $\pm$ 0.02 & 0.07 $\pm$ 0.02 & 0.31 $\pm$ 0.06\\
		$kT_{2}$ (keV) & - & 0.75 $\pm$ 0.097 & 1.29 $\pm$ 0.13\\
		%Normalisation 1 & 0.05 $\pm$  0.06 &  & 5.7 $\pm$ 1.5 ($\times10^{-5}$)\\
		%Normalisation 2 & - &  & 10.7 $\pm$  2.92 ($\times10^{-5}$)\\
		$\chi^{2}_{\mathrm{red}}$ & 1.42 & 1.43 & 1.27\\
		
		$N_{\textsc{h}}$ ($10^{22}$ cm$^{-2}$) & 0.83 $\pm$ 0.11 & 0.37 $\pm$ 0.14 & -\\
		$F_{\mathrm{X}}$ (erg cm$^{-2}$ s$^{-1}$) &  2.8$\times10^{-13}$  & 3.3$\times10^{-13}$ & 3.4$\times10^{-13}$\\		
		
		\hline
	\end{tabular}
\end{table}
%===================================================

%===================================================
% Example table
\begin{table*}
\begin{minipage}{155mm}

	\centering
	\caption{Flux and luminosity in X-ray and UV. Luminosity is calculated using the distance of 113.4 pc from Gaia DR2. X-ray luminosity is based on the flux from the 2-temperature \textsc{apec}  model without absorption. Flux density is the flux at the central wavelength calculated by the \textsc{uvotflux} software. This has been converted to a flux for the filter range by assuming the flux density is equal at all wavelengths and weighting by the filter transmission function before integrating. Flux at planet values do not include absorption by the disc.}
	\label{Table:table_4_flux_lum}
	\begin{tabular}{lccccc} % four columns, alignment for each
		\hline
		 & X-ray (0.3-10.0 keV) & uvw2 & uvw1 & U & Units\\
		Range & (1.2 - 41 \AA) & (1599-2256 \AA) & (2253-2946 \AA) &(3072-3857 \AA) &\\
		Central wavelength &  & 1928 \AA & 2600 \AA & 3465 \AA &\\
		\hline
		Flux density & - & 1.5$\times 10^{-15}$ & 2.4$\times 10^{-15}$ & 2.2$\times 10^{-15}$ & erg cm$^{-2}$ s$^{-1}$ \AA$^{-1}$\\ 
		Flux & 3.4$\times 10^{-13}$ &  1.5$\times 10^{-13}$ & 2.7$\times 10^{-13}$ & 3.5$\times 10^{-13}$ & erg cm$^{-2}$ s$^{-1}$\\
		Luminosity & 5.2$\times 10^{29}$ & 2.3$\times 10^{29}$ & 4.2$\times 10^{29}$ & 5.4$\times 10^{29}$ & erg s$^{-1}$\\
		log $L_\mathrm{x}/L_\mathrm{bol}$ & -3.41 &  & & &\\
		Flux at PDS 70b & 0.43 & 0.19 & 0.35 & 0.45 & erg cm$^{-2}$ s$^{-1}$\\
		Flux at PDS 70c  & 0.16 & 0.07 & 0.12 & 0.16 & erg cm$^{-2}$ s$^{-1}$\\
		\hline
	\end{tabular}

	\end{minipage}
\end{table*}
%===================================================

\section{Discussion}

The UV emission can be used to measure the accretion rate of material from the disc on to the central star. As material impacts the star, the release of kinetic energy heats plasma to temperatures of $\sim$ 10,000 K which emits UV radiation. UV emission from accretion dominates the stellar photospheric emission in the 2000-3500 \AA\ range (See Fig 2b of \citealt{Hartmann_2016}). Accretion emission should be observable in the 3406-3493 \AA\ and 2475-2545 \AA\ wavelength range covered by the UVOT U and uvw1 filter. 

It is also expected that there is a contribution to the UV flux due to chromospheric activity. As a first approximation for an upper limit of the accretion rate we assume the observed UV flux is entirely produced by the accretion shock. The accretion luminosity is related to the stellar mass, radius and the accretion rate by equation 1 \citep{Venuti_2014, Gullbring_1998}, where $R_{\mathrm{in}}$ is the inner edge of the accretion disc and is taken as $R_{\textrm{in}} = 5\times R_{*}$.

\begin{equation}\label{Eqn:accretion_rate}
\dot{M}_{\mathrm{acc}} = \left(1-\frac{R_{*}}{R_{\mathrm{in}}}\right)^{-1}\frac{L_{\mathrm{acc}}R_{*}}{GM_{*}}
\end{equation}

The UVOT U band filter closely corresponds to the range of the U filter used in \cite{Venuti_2014} which had a FWHM range of 3110-3970 \AA . Using the stellar parameters in Table \ref{Table:table_2_stellar_properties} and the U band luminosity of 5.4$\times 10^{29}$ erg/s gives a mass accretion rate of 6.0$\times 10^{-12}$ $M_{\odot}$/yr. This is much lower than typical accretion rates for T-Tauri stars such as those in the NGC 2264 cluster which range from $M_{\mathrm{acc}} = 1\times 10^{-10}$ up to $1\times 10^{-7}$ $M_{\odot}$/yr \citep{Venuti_2014}. This low accretion rate indicates that PDS 70 is not currently accreting material from the disc and the residual UV flux is more likely to be due to magnetic activity in the stellar chromosphere. This is consistent with previous studies which classified PDS 70 as a weak-line T-Tauri star showing no evidence of accretion in H-$\alpha$ observations \citep{Gregorio_Hetem_2002}. Comparison with the luminosities observed for known accreting stars (Classical T-Tauris) of similar spectral type in \cite{Gullbring_1998} shows they have accretion luminosities (UV) from 0.007 to 0.6 $L_{\odot}$, while PDS 70 has a UV luminosity of only 0.0004 $L_{\odot}$. It is possible that the low UV luminosity is partly due to the accretion shock region being obscured by the disc. However the disc inclination of 50 degrees (far from edge-on) makes this unlikely.

The \textit{ROSAT} All-Sky Survey in 1990 \citep{Boller_2016} detected a source which was identified as an X-ray counterpart to PDS 70 by \cite{Haakonsen_2009} and \cite{Kiraga_2012}. The X-ray flux observed by \textit{Swift} is slightly lower, having decreased from $\sim$5-8$\times 10^{-13}$ in 1990 to 3.4$\times 10^{-13}$ erg cm$^{-2}$ s$^{-1}$ in 2019, albeit the \textit{ROSAT} measurements were in a somewhat lower photon-energy band 0.1-2.4 keV. 

The planets are quite far from the star (20 and 35 AU) so there is unlikely to be any significant effect from the stellar radiation such as atmospheric evaporation. The absorption by the disc between star and planet is unknown and is the main uncertainty in estimating the irradiation of the planets. Even though the planets are in a gap, the thickness of the  disc either side is likely to be $\sim$2 AU in height so could still shield the planet from the star's direct radiation \citep{Dong_2012}. We calculate upper limits on the irradiation of the planet assuming no attenuation of XUV by the disc. These are calculated for the X-ray and UV bands shown in Table \ref{Table:table_4_flux_lum}. The upper limit on the X-ray flux at planet b and c is 0.43 and 0.16 erg cm$^{-2}$ s$^{-1}$ respectively. This is less than that received by the present day Earth which is 0.85 erg cm$^{-2}$ s$^{-1}$ \citep{Ribas_2005}.

\section{Conclusions}

PDS 70 is detected in both X-rays and UV. The target is detected in UV for the first time and is visible in all 3 UV wavebands observed. The X-ray and UV light-curves show very little variability in 2 ks snapshots spread over 5 weeks except for a slightly higher UV flux ($\sim2\times 10^{-13}$ erg cm$^{-2}$ s$^{-1}$ above average) in the first observation. An X-ray source is clearly detected when all observations are combined to make a total exposure of 10 ks. The spectrum is best modelled by a 2-temperature plasma with no significant absorption by interstellar neutral hydrogen. The plasma components are 0.31 keV and 1.29 keV which is similar to results found for LkCa 15 which is a similar age and spectral type.  

The X-ray flux (0.3-10.0 keV) derived from the best fit model is 3.4$\times 10^{-13}$ erg cm$^{-2}$ s$^{-1}$ and the luminosity using 113.4 pc distance is 5.2$\times 10^{29}$ erg s$^{-1}$. This is typical of X-ray luminosities seen in other T-Tauri stars. The most likely source of the X-ray emission is stellar coronal activity which is consistent with the emission peak at 1 keV. The log $\frac{L_{\mathrm{X}}}{L_{\mathrm{bol}}}$ ratio of -3.4 is consistent with observations of young, rapidly rotating stars near the saturation limit.

The UV luminosity is much lower than would be expected if it was produced by an accretion shock from disc material accreting onto the star. The maximum accretion rate, given the observed UV luminosity, is 6.0$\times 10^{-12}$ $M_{\odot}$/yr, which is significantly lower than the 1$\times 10^{-10}$ or greater accretion rate estimated for known accreting T-Tauri stars. This indicates that no significant accretion from the disc to the star is currently taking place in the PDS 70 system and the UV flux is more likely coronal in origin. The planets are likely to be shielded from XUV irradiation by the disc, but even neglecting absorption, the XUV flux at the distance of planet b and c is lower than that received by the present day Earth. The stellar XUV radiation, is only likely to significantly affect the planets if they migrate inward to close ($\sim$0.1 AU) orbits like other hot-Jupiters. 

By contrast, the effects of the observed stellar XUV flux on the disc are expected to be significant. Models of X-ray photoevaporation show that an X-ray luminosity of $L_X = 5\times10^{29}$\,erg\,s$^{-1}$ produces a wind mass-loss rate of $\dot{M}_{\mathrm w}\simeq 10^{-8}$\,M$_{\odot}$\,yr$^{-1}$ \citep{Owen_2010,Owen_2012,Picogna_2019}. The mass of gas in the PDS70 disc is not well-constrained, but for standard assumptions (i.e., a gas-to-dust ratio of 100) the disc models used to fit the dust observations imply a total disc mass $M_{\mathrm d} \lesssim 10^{-2}$--$10^{-3}$M$_{\odot}$ (e.g. \citealt{Keppler_2018}). This in turn implies a relatively short disc dispersal time ($M_{\mathrm d}/\dot{M}_{\mathrm w}<1$Myr), significantly less than the current $\simeq 5$\,Myr age of the system. The result is that photevoporation, rather than planet formation or accretion, is expected to dominate the future evolution of the disc. Moreover, the lack of accretion on to the star suggests that there is a cavity in the gas as well as in the dust, in agreement with IR observations by \cite{Long_2018}. In such a configuration, direct observational tracers of disc photoevaporation (such as blue-shifted emission lines) should be detectable \citep[e.g.,][]{Ercolano_2010,Ercolano_2016}. The PDS70 system may therefore represent a critical benchmark for our understanding of protoplanetary disc evolution.

\section*{Acknowledgements}

We thank the referee for helpful comments. Special thanks to Klaas Wiersema, Nial Tanvir and Laura Venuti for illuminating discussions, and to Thomas Henning (HIFOL) for enthusiastically introducing us to PDS 70. This work was carried out as part of the ExoplANETS-A project http://exoplanet-atmosphere.eu (Exoplanet Atmosphere New Emission Transmission Spectra Analysis);
https://cordis.europa.eu/project/rcn/212911 en.html; The ExoplANETS-A project is funded from the EU's Horizon-2020 programme; Grant Agreement no. 776403. This work used data supplied by the UK Swift Science Data Centre, University of Leicester. KLP ackowledges support from the UK Space Agency. We acknowledge use of CDS SIMBAD and NASA ADS facilities. RA received funding from the European Research Council (grant agreement No 681601).

%%%%%%%%%%%%%%%%%%%%%%%%%%%%%%%%%%%%%%%%%%%%%%%%%%

%%%%%%%%%%%%%%%%%%%% REFERENCES %%%%%%%%%%%%%%%%%%

% The best way to enter references is to use BibTeX:

%\bibliographystyle{mnras}
%\bibliography{example} % if your bibtex file is called example.bib

\begin{thebibliography}{99}
%\bibitem[\protect\citeauthoryear{Author}{2012}]{Author2012}
%Author A.~N., 2013, Journal of Improbable Astronomy, 1, 1
%\bibitem[\protect\citeauthoryear{Others}{2013}]{Others2013}
%Others S., 2012, Journal of Interesting Stuff, 17, 198
%============

\bibitem[\protect\citeauthoryear{Alexander, Clarke \& Pringle}{2006}]{Alexander_2006} Alexander R.~D., Clarke C.~J., Pringle J.~E., 2006, MNRAS, 369, 216


\bibitem[\protect\citeauthoryear{Alexander et al.}{2014}]{Alexander_2014} Alexander R., Pascucci I., Andrews S., Armitage P., Cieza L., 2014, prpl.conf,  475, prpl.conf


\bibitem[\protect\citeauthoryear{Boller, et al.}{2016}]{Boller_2016} Boller T., Freyberg M.~J., Tr{\"u}mper J., Haberl F., Voges W., Nandra K., 2016, A\&A, 588, A103


\bibitem[\protect\citeauthoryear{Christiaens et al.}{2019a}]{Christiaens_2019_ApJ_a} Christiaens V., Cantalloube F., Casassus S., Price D.~J., Absil O., Pinte C., Girard J., Montesinos M., 2019, ApJ, 877, L33 


\bibitem[\protect\citeauthoryear{Christiaens et al.}{2019b}]{Christiaens_2019_MNRAS_b} Christiaens V., et al., 2019, MNRAS, 486, 5819 


\bibitem[\protect\citeauthoryear{Dong et al.}{2012}]{Dong_2012} Dong R., et al., 2012, ApJ, 760, 111 

\bibitem[\protect\citeauthoryear{Ercolano \& Owen}{2010}]{Ercolano_2010} Ercolano B., Owen J.~E., 2010, MNRAS, 406, 1553

\bibitem[\protect\citeauthoryear{Ercolano \& Owen}{2016}]{Ercolano_2016} Ercolano B., Owen J.~E., 2016, MNRAS, 460, 3472

\bibitem[\protect\citeauthoryear{Ercolano \& Pascucci}{2017}]{Ercolano_Pascucci_2017} Ercolano B., Pascucci I., 2017, RSOS, 4, 170114

\bibitem[Evans et al.(2007)]{Evans_07} Evans, P.~A., Beardmore, A.~P., Page, K.~L., et al.\ 2007, A\&A, 469, 379 

\bibitem[\protect\citeauthoryear{Evans, et al.}{2009}]{Evans_09} Evans P.~A., et al., 2009, MNRAS, 397, 1177

\bibitem[\protect\citeauthoryear{Gehrels, et al.}{2004}]{Gehrels_2004} Gehrels N., et al., 2004, ApJ, 611, 1005

\bibitem[\protect\citeauthoryear{Gorti et al.}{2016}]{Gorti_2016} Gorti U., Liseau R., S{\'a}ndor Z., Clarke C., 2016, SSRv, 205, 125

\bibitem[\protect\citeauthoryear{Gorti \& Hollenbach}{2009}]{Gorti_Hollenbach_2009} Gorti U., Hollenbach D., 2009, ApJ, 690, 1539

\bibitem[\protect\citeauthoryear{Gregorio-Hetem \& Hetem}{2002}]{Gregorio_Hetem_2002} Gregorio-Hetem J., Hetem A., 2002, MNRAS, 336, 197

\bibitem[\protect\citeauthoryear{G{\"u}del \& Naz{\'e}}{2009}]{Gudel_Naze_2009} G{\"u}del M., Naz{\'e} Y., 2009, A\&ARv, 17, 309


\bibitem[\protect\citeauthoryear{Gullbring, et al.}{1998}]{Gullbring_1998} Gullbring E., Hartmann L., Brice{\~n}o C., Calvet N., 1998, ApJ, 492, 323

\bibitem[\protect\citeauthoryear{Haakonsen \& Rutledge}{2009}]{Haakonsen_2009} Haakonsen C.~B., Rutledge R.~E., 2009, ApJS, 184, 138


\bibitem[\protect\citeauthoryear{Haffert et al.}{2019}]{Haffert_2019} Haffert S.~Y., Bohn A.~J., de Boer J., Snellen I.~A.~G., Brinchmann J., Girard J.~H., Keller C.~U., Bacon R., 2019, NatAs 

\bibitem[\protect\citeauthoryear{Haisch, Lada \& Lada}{2001}]{Haisch_Lada_Lada_2001} Haisch K.~E., Lada E.~A., Lada C.~J., 2001, ApJL, 553, L153 

\bibitem[\protect\citeauthoryear{Hartmann, Herczeg \& Calvet}{2016}]{Hartmann_2016} Hartmann L., Herczeg G., Calvet N., 2016, ARA\&A, 54, 135

\bibitem[\protect\citeauthoryear{Hashimoto et al.}{2012}]{Hashimoto_2012} Hashimoto J., et al., 2012, ApJ, 758, L19 


\bibitem[\protect\citeauthoryear{Hashimoto et al.}{2015}]{Hashimoto_2015} Hashimoto J., et al., 2015, ApJ, 799, 43 


\bibitem[\protect\citeauthoryear{Isella, et al.}{2019}]{Isella_2019} Isella A., Benisty M., Teague R., Bae J., Keppler M., Facchini S., P{\'e}rez L., 2019, ApJL, 879, L25

\bibitem[\protect\citeauthoryear{Jackson, Davis \& Wheatley}{2012}]{Jackson_Davis_Wheatley_2012} Jackson A.~P., Davis T.~A., Wheatley P.~J., 2012, MNRAS, 422, 2024

\bibitem[\protect\citeauthoryear{Keppler et al.}{2018}]{Keppler_2018} Keppler M., et al., 2018, A\&A, 617, A44

\bibitem[\protect\citeauthoryear{Keppler et al.}{2019}]{Keppler_2019} Keppler M., et al., 2019, A\&A, 625, A118 

\bibitem[\protect\citeauthoryear{Kiraga}{2012}]{Kiraga_2012} Kiraga M., 2012, AcA, 62, 67

\bibitem[\protect\citeauthoryear{Kley \& Nelson}{2012}]{Kley_Nelson_2012} Kley W., Nelson R.~P., 2012, ARA\&A, 50, 211

\bibitem[\protect\citeauthoryear{Long et al.}{2018}]{Long_2018} Long Z.~C., et al., 2018, ApJ, 858, 112 

\bibitem[\protect\citeauthoryear{Lunine \& Stevenson}{1982}]{Lunine_Stevenson_1982} Lunine J.~I., Stevenson D.~J., 1982, Icar, 52, 14

\bibitem[\protect\citeauthoryear{Mamajek}{2009}]{Mamajek_2009} Mamajek E.~E., 2009, AIPC,  3, AIPC.1158

\bibitem[\protect\citeauthoryear{M{\"u}ller, et al.}{2018}]{Muller_2018} M{\"u}ller A., et al., 2018, A\&A, 617, L2


\bibitem[\protect\citeauthoryear{O'dell, Wen \& Hu}{1993}]{Odell_1993} O'dell C.~R., Wen Z., Hu X., 1993, ApJ, 410, 696

\bibitem[\protect\citeauthoryear{Owen, et al.}{2010}]{Owen_2010} Owen J.~E., Ercolano B., Clarke C.~J., Alexander R.~D., 2010, MNRAS, 401, 1415

\bibitem[\protect\citeauthoryear{Owen \& Jackson}{2012}]{Owen_2012} Owen J.~E., Jackson A.~P., 2012, MNRAS, 425, 2931

\bibitem[\protect\citeauthoryear{Pallavicini, Golub, Rosner et al.}{1981}]{Pallavicini_1981} Pallavicini R., Golub L., Rosner R., Vaiana G.~S., Ayres T., Linsky J.~L., 1981, ApJ, 248, 279


\bibitem[\protect\citeauthoryear{Pascucci et al.}{2014}]{Pascucci_2014} Pascucci I., Ricci L., Gorti U., et al., 2014, ApJ, 795, 1

\bibitem[\protect\citeauthoryear{Pecaut \& Mamajek}{2016}]{Pecaut_Mamajek_2016} Pecaut M.~J., Mamajek E.~E., 2016, MNRAS, 461, 794

\bibitem[\protect\citeauthoryear{Picogna, et al.}{2019}]{Picogna_2019} Picogna G., Ercolano B., Owen J.~E., Weber M.~L., 2019, MNRAS, 487, 691

\bibitem[\protect\citeauthoryear{Pizzolato, Maggio, Micela et al.}{2003}]{Pizzolato_2003} Pizzolato N., Maggio A., Micela G., et al., 2003, A\&A, 397, 147


\bibitem[\protect\citeauthoryear{Riaud et al.}{2006}]{Riaud_2006} Riaud P., Mawet D., Absil O., Boccaletti A., Baudoz P., Herwats E., Surdej J., 2006, A\&A, 458, 317 

\bibitem[\protect\citeauthoryear{Ribas, Bouy \& Mer{\'\i}n}{2015}]{Ribas_2015} Ribas {\'A}., Bouy H., Mer{\'\i}n B., 2015, A\&A, 576, A52

\bibitem[\protect\citeauthoryear{Ribas, et al.}{2005}]{Ribas_2005} Ribas I., Guinan E.~F., G{\"u}del M., et al., 2005, ApJ, 622, 680

\bibitem[\protect\citeauthoryear{Skinner \& G{\"u}del}{2013}]{Skinner_Gudel_2013} Skinner S.~L., G{\"u}del M., 2013, ApJ, 765, 3

\bibitem[\protect\citeauthoryear{Smith, et al.}{2001}]{Smith_2001} Smith R.~K., Brickhouse N.~S., Liedahl D.~A., Raymond J.~C., 2001, ApJL, 556, L91

\bibitem[\protect\citeauthoryear{Venuti, et al.}{2014}]{Venuti_2014} Venuti L., et al., 2014, A\&A, 570, A82

\bibitem[\protect\citeauthoryear{Wagner et al.}{2018}]{Wagner_2018} Wagner K., et al., 2018, ApJ, 863, L8 

\bibitem[\protect\citeauthoryear{Wheatley et al.}{2017}]{Wheatley_2017} Wheatley P.~J., Louden T., Bourrier V., et al., 2017, MNRAS, 465, L74


\bibitem[\protect\citeauthoryear{Zhu et al.}{2011}]{Zhu_2011} Zhu Z., et al, 2011, ApJ, 729, 47

\bibitem[\protect\citeauthoryear{Zuckerman, Forveille \& Kastner}{1995}]{Zuckerman_1995} Zuckerman B., Forveille T., Kastner J.~H., 1995, Natur, 373, 494

%============
\end{thebibliography}

% Alternatively you could enter them by hand, like this:
% This method is tedious and prone to error if you have lots of references

%%%%%%%%%%%%%%%%%%%%%%%%%%%%%%%%%%%%%%%%%%%%%%%%%%

%%%%%%%%%%%%%%%%% APPENDICES %%%%%%%%%%%%%%%%%%%%%

\appendix

%\section{Some extra material}

%If you want to present additional material which would interrupt the flow of the main paper,
%it can be placed in an Appendix which appears after the list of references.

%%%%%%%%%%%%%%%%%%%%%%%%%%%%%%%%%%%%%%%%%%%%%%%%%%

% Don't change these lines
\bsp	% typesetting comment
\label{lastpage}
\end{document}